\documentstyle[amstex,art8,aas2pp4,flushrt,tighten,xr]{article}

\newcommand{\pa}{\partial}

\newcommand{\cri}{_{\rm c}}
\newcommand{\zm}{z_{\rm m}}
\newcommand{\zM}{z_{\rm M}}

\received{}
\accepted{}
\journalid{}{}
\articleid{}{}

\slugcomment{Submitted to the Astrophysical Journal}

\begin{document}

\title{RUNAWAY ACCELERATION OF LINE DRIVEN WINDS: THE ROLE OF THE
       OUTER BOUNDARY}

\author{Achim Feldmeier}

\affil{Astrophysik, Institut f\"ur Physik, Universit\"at Potsdam, Am
Neuen Palais 10, 14469 Potsdam, Germany, E-mail: {\tt
afeld@@astro.physik.uni-potsdam.de}}

\and

\author{Isaac Shlosman\altaffilmark{1,2}}

\affil{Joint Institute for Laboratory Astrophysics, University of Colorado,
    Box 440, Boulder, CO 80309-440, USA, E-mail: {\tt shlosman@@pa.uky.edu}}

\altaffiltext{1}{JILA Visiting Fellow}
\altaffiltext{2}{permanent address: Department of Physics and Astronomy,
University of Kentucky, Lexington, KY 40506-0055} 

\and

\author{Wolf-Rainer Hamann}

\affil{Astrophysik, Institut f\"ur Physik, Universit\"at Potsdam, Am
Neuen Palais 10, 14469 Potsdam, Germany, E-mail: {\tt
wrh@@astro.physik.uni-potsdam.de}}

\begin{abstract}

Observations and theory suggest that line driven winds from hot stars
and luminous accretion disks adopt a unique, critical solution which
corresponds to maximum mass loss rate. We analyze the numerical
stability of the infinite family of shallow wind solutions, which
resemble solar wind breezes, and their transition to the critical
wind. Shallow solutions are sub-critical with respect to radiative (or
Abbott) waves. These waves can propagate upstream through shallow
winds at high speeds. If the waves are not accounted for in the
Courant time step, numerical runaway results. The outer boundary
condition is equally important for wind stability. Assuming pure
outflow conditions, as is done in the literature, triggers runaway of
shallow winds to the critical solution or to accretion flow.

\end{abstract}


\section{Introduction}
\label{introduction}

Line driven winds (LDWs) occur in various astronomical objects, like
OB and Wolf-Rayet stars, in accretion disks in cataclysmic variables
and, probably, in active galactic nuclei and luminous young stellar
objects. These winds are powered by absorption and re-emission of
photospheric continuum flux in numerous spectral transitions of C, N,
O, Fe, etc.  ions.

Castor, Abbott, \& Klein (1975; hereafter CAK) have analyzed the
steady state Euler equation for LDWs. They found an infinite family of
mathematical solutions, but only one, hereafter `critical solution',
which extends from the photosphere to arbitrary large radii. Other
solutions either do not reach infinity or the photosphere. The former
solutions are called shallow and the latter ones steep. The unique,
critical wind starts as the fastest shallow solution and switches
smoothly to the slowest steep solution at the critical point.

Observational support that LDWs adopt the critical solution comes from
measured terminal speeds (Abbott 1982). Furthermore, mass loss rates
of supergiant winds are in general agreement with modified CAK theory
(Lamers \& Leitherer 1993; Puls et al.~1996). These measurements were
recently extended to include galactic and extragalactic OB and A stars
and central stars of planetary nebula (Kudritzki et al.~1999).

Abbott (1980) put CAK theory in a complete analogy to the solar wind
and nozzle flows. The existence of a sonic point defines the unique,
transsonic solutions for these flows, whereas the existence of a
critical point for Abbott waves defines the unique, CAK solution for
LDWs. Only from below this critical point, Abbott waves can propagate
upstream towards the photosphere. Above the critical point, they are
advected outwards. Because Abbott waves generally propagate highly
supersonically, the critical point of LDWs lies at much higher speeds
than the sonic point.

Abbott's (1980) analysis was challenged by Owocki \& Rybicki (1986),
who derived the Green's function for a pure absorption LDW. The
Green's function gives correct signal speeds in presence of
hydrodynamic instabilities. The inward signal speed in a pure
absorption line wind is the sound speed, and not the much larger
Abbott speed, because photons propagate outwards only. Owocki \&
Rybicki (1986) showed that a fiducial upstream signal, which still
propagates inward at Abbott speed, must be interpreted as {\it purely
local} Taylor series reconstruction. For a flow driven by scattering
lines, however, Owocki \& Puls (1999) find {\it physically relevant}
Abbott waves for a numerical Green's function.

In the present paper, we further analyze the properties of Abbott
waves. We show that they are crucial for our understanding of
stability of LDWs and must be included in the Courant time step. So
far, time-dependent numerical simulations of LDWs from stars and
accretion disks have ignored the ability of Abbott waves to
communicate in the supersonic regime, which results in a numerical
runaway. In particular, this runaway can lift the wind to the critical
solution.

The critical solution is also enforced by applying pure outflow
boundary conditions. It is often argued that outflow boundary
conditions are appropriate since LDWs are highly supersonic. Instead,
they have to be {\it super-abbottic}. We show that shallow wind
solutions, which correspond to solar wind breezes, are everywhere
sub-abbottic. Hence, these solutions are numerically destabilized by
applying outflow boundary conditions.

We formulate boundary conditions which render shallow solutions
numerically stable. Those include non-reflecting Riemann conditions
for Abbott waves. By allowing for kinks in the velocity law, shallow
solutions can be made globally admissible.

\section{The wind model}
\label{windmodel}

In the CAK model for LDWs, both gravity and line force scale with
$r^{-2}$. If the sound speed and hence the pressure forces are set to
zero, this leads to a degeneracy of the critical point condition,
which is satisfied formally at every radius (Poe, Owocki, \&
Castor~1990). Thus, for this case, Abbott waves cannot propagate
inwards from any location in the wind. For finite sound speed, they
creep inwards at small speed. Inclusion of the finite disk correction
factor is much more relevant for LDWs than inclusion of pressure
forces. With the finite disk included, the inward speed of Abbott
waves below the critical point is significantly larger than the wind
speed.  Unfortunately, the finite disk correction factor depends on
the (unknown) velocity law of the wind, which prevents a simple
analysis of the wind dynamics.

We consider, therefore, a wind model which is analytically feasible
and yet prevents the (near-)degeneracy of the CAK point-star wind.
(Especially, the latter leads to poor convergence of time-dependent
numerical schemes.) As a prototype, a vertical LDW from an isothermal, 
geometrically thin, non-self-gravitating accretion disk is assumed. The sound
speed is set to zero. Keplerian rotation is assumed within the disk and angular
momentum conservation above the disk. This reduces the flow problem to
a 1-D, planar one. The radiative flux above an isothermal disk is
roughly constant at small heights. On the other hand, the vertical
gravity component along the wind cylinder is zero in the disk
midplane, grows linearly with $z$ if $z\ll R$ (with $R$ the footpoint
radius in the disk), reaches a maximum, and drops off at large $z$. To
model the launch region of the wind and the gravity maximum, we choose
$g(z)=z/(1+z^2)$, with normalization $GM=1$ and $R=1$, $G$ being
gravitational constant, and $M$ is the mass of the central object.
The different spatial dependence of flux and gravity results in a
well-defined critical point in the flow.

For constant radiative flux, the CAK line force becomes $g_{\rm l} = C
(\rho^{-1} dv/dz)^\alpha$, where $\rho$ and $v$ are the density and
velocity, and $\alpha$ and $C$ are constants. We choose $\alpha=1/2$.
Typical values of $\alpha$ from NLTE calculations (Pauldrach 1987) and
Kramers' law (Puls, Springmann, \& Lennon 2000) are $\alpha=0.4$ to
0.7. The constant $C$ can be expressed in terms of the unique,
critical CAK mass loss rate. This is the maximum allowed mass loss
rate for a stationary wind. We introduce new fluid variables $m$ and
$w$,
 \begin{equation}
 \label{eq2}
 m = {\rho v \over \rho\cri v\cri},\qquad\quad
 w = {1\over 2} v^2,
 \end{equation}
 where a subsript `c' refers to the critical CAK solution. For
stationary, plane parallel winds, the continuity equation becomes
$\rho v= {\rm const}$, and $m$ is a constant. In the following, the quantity
$w'=dw/dz=v\,dv/dz$ will play a central role. The velocity law, $v(z)$, is
obtained by quadrature of the wind `solution' $w'(z)$. For $\alpha=1/2$, the
stationary Euler equation becomes,
 \begin{equation}
 \label{euler}
 E = w'+g(z)- 2\sqrt{g\cri w'/m} = 0,
 \end{equation}
 where we multiplied the nominator and denominator in the line force
by $v$, and introduced $g\cri=g(z\cri)$, the gravity at the location
of the CAK critical point. The dependence of $w'$ on $z$ is supressed
in the equation. The constant $C$ was determined as follows.
Equation~(\ref{euler}) is a quadratic equation in $\sqrt{w'}$.
Consider the $\sqrt{w'}E$ plane. For arbitrary $z$ and sufficiently
small $m$, the crossings $E=0$ of the parabola $E(\sqrt{w'})$ with the
abscissa determine two wind solutions $w'(z)$, termed shallow and
steep. When $m$ increases, the crossings approach each other, and
merge at some maximum $m\ge 1$, beyond which no further solutions
exist. Hence, the {\it singularity condition} $\pa E/\pa w'=0$ holds
when two solutions merge, as is the case at the critical point.
Together with $E=0$, this gives $w'\cri=g\cri$ and
$C=2\sqrt{g\cri\rho\cri v\cri}$.
 
 We add a comment which is important for the following. At every
location $z\ne z\cri$, $m>1$ for the parabola which meets the abscissa
in just one point. The minimum, $m=1$, is reached at the unique
critical point, $z\cri$, which is the flow {\it nozzle} and determines
the maximum mass loss which can be driven from the photosphere to
infinity. Solutions (actually, pairs of shallow and steep solutions)
with $m>1$ are called {\it overloaded}. They correspond to a {\it
choked} Laval nozzle flow. These solutions become imaginary in a
neigborhood of the critical point. We have shown elsewhere (Feldmeier
\& Shlosman 2000) how overloaded LDW solutions can be made globally
defined, by letting them decelerate, $dv/dz<0$, around the critical
point.

The {\it position} of the critical point is determined by the {\it
regularity} condition, which is derived next. Since $E=0$ everywhere,
$dE/dz=0$ holds too. Since $\pa E/\pa w'\cri=0$, it follows that $\pa
E/\pa z\cri=0$, if $|w''\cri|<\infty$. This latter condition singles 
out the unique critical point, since overloaded winds have kinks at the
merging point of two solutions, $|w''|\equiv\infty$.
Hence, the critical point lies at the gravity maximum,
$dg/dz\cri=0$. For $g(z)=z/(1+z^2)$, $z\cri=1$ and $g\cri =1/2$. The
line force in (\ref{euler}) becomes $\sqrt{2w'/m}$. This holds for any
solution, whether shallow or steep. Solving (\ref{euler}) for
$\sqrt{w'}$,
 \begin{equation}
 \label{eulersol}
 q(z)\equiv \sqrt{2mw'(z)} = 1\pm\sqrt{1-2mg(z)},
 \end{equation}
 where we introduced a new variable $q$. The signs refer to steep
($+$) and shallow ($-$) solutions. For $m=1$, the square root in
(\ref{eulersol}) vanishes at the critical point. For $m>1$, the root
becomes imaginary in a neighborhood of the critical point. The
solutions $w'(z)$ of (\ref{eulersol}) have a saddle topology in the
$zw'$ plane when $m$ is varied from $m<1$ to $m>1$. The critical point
$z\cri=1, w'\cri=1/2$ is the saddle point. A variable radiative flux
above a non-isothermal accretion disk leads instead to two different
$z\cri$, above and below the gravity maximum (Feldmeier \& Shlosman
1999). The lower is a saddle, the upper an extremum.

Note that the CAK critical point of LDWs is a saddle in the $z-vv'$
plane, whereas the sonic point of the solar wind is a saddle in the
$zv$ plane. In all other respects, we find a deep analogy between LDWs
and the solar wind. Only the notion of a critical speed is replaced by
a critical acceleration. Shallow solutions are the analogs to solar
wind breezes. 

According to (\ref{eulersol}), shallow and steep solutions with $m\le
1$ extend from $z=0$ to $\infty$ for the present model with zero sound
speed. CAK found that steep solutions start supersonically at the wind
base $z=0$, which seems unphysical. In a spherically symmetric wind,
shallow solutions do not extend to arbitrarily large $z$, because they
cannot provide the required expansion work. These two results and the
requirement for a continuous and differentiable solution implies that
the wind has to pass through the critical point. It starts on the
fastest shallow, and ends on the slowest steep solution.

Figure~1 (which is adapted from Fig.~4 in Abbott 1980 or
Fig.~8.13 in Lamers \& Cassinelli 1999) shows the LDW solution
topology in the $rv$ plane, for a finite sound speed. In region (II),
which is spatially the most extended in real LDWs, shallow and steep
solutions exist at each point $(r,v)$. If $z\ne z\cri$, these solution
pairs may be overloaded. In the subsonic region (I), only shallow
solutions (possibly overloaded) exist; steep solutions are necessarily
supersonic. Correspondingly, only steep solutions (possibly
overloaded) exist in region (III). No solutions exist in regions (IV)
and (V). Note that the gap in overloaded solutions around $z\cri$ can
be bridged by an extended decelerating region, $dv/dr<0$.

 \begin{figure}[ht]
 \begin{center}
 \leavevmode
 \epsfxsize=8.4cm
 \epsffile{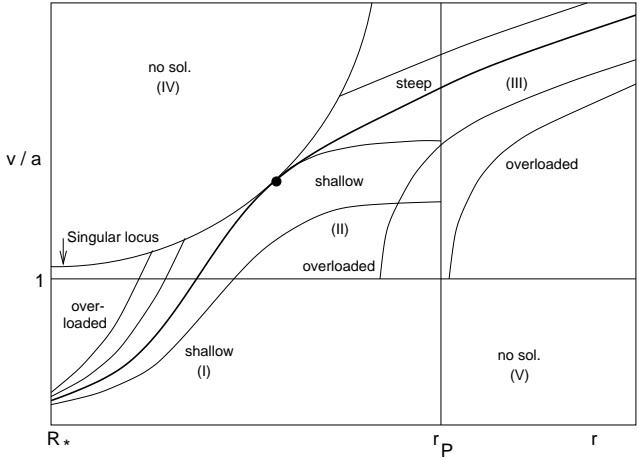}  
 \caption{Solution topology of line driven winds, showing
shallow, steep, and overloaded solutions. The bullet marks the CAK
critical point. Roman numerals refer to cases (I) to (V) of CAK.
Regions (I,II,III), and (V) are mutually separated by straight lines
(sound speed $v=a$; Parker point $r=r_P$); regions (II) and (IV) are
separated by the singular locus.\label{fig1}}
 \end{center}
 \end{figure}  

The argument by CAK ruling out global, steep solutions is stringent,
but the exclusion of shallow solutions seems too restrictive. The
pressure force due to spherical expansion becomes important only at
very large distances from the wind base. According to CAK, shallow solutions
break down around 300 stellar radii for realistic O~star wind parameters. Most
shallow solutions have by then overcome the local escape speed.
Shallow solutions can be globally defined by a slight generalization
of the CAK approach, allowing for negative $v'=dv/dz$ in the line
force $\sim \sqrt{v'}$. This can be done by replacing $v'$ by $|v'|$
or by $\max(v',0)$. The former expresses the `blindness' of purely
local line driving to flow acceleration or deceleration. The latter
accounts for the nonlocal effect of shadowing in non-monotonic
velocity laws, creating a resonance coupling. All radiation is assumed
to be absorbed at the first resonance location. Realistically, the
value of the line force lies between these extremes. The decelerating
wind solution, $w'<0$, is unique, and does not split into shallow and
steep branches.

Hyperbolic differential equations allow for weak discontinuities,
which are represented in the $zw'$ plane by jumps between different
solution branches. Hence, a shallow solution may jump onto the
decelerating branch. Even an overloaded solution can avoid becoming
imaginary by jumping onto the decelerating branch in the vicinity of
the critical point. This opens the possibility that global solutions
can be composed in a piecewise manner.

Jumps in $w'$ introduce {\it kinks} in $v(z)$. Such kinks have been
found on various occasions in time-dependent LDW simulations. They may
play a central role in the observed {\it discrete absorption
components} in non-saturated P~Cygni line profiles (Cranmer \& Owocki
1996). In LDWs from accretion disks in cataclysmic variables, a jump
onto the decelerating branch even occurs for the critical solution,
which would elsewhere not extend to infinity (Feldmeier \& Shlosman
1999). Hence, the smoothness argument of CAK cannot be used to
disqualify shallow solutions.

\section{Abbott waves}
\label{abbottwaves}

\subsection{Linear dispersion analysis}

The time-dependent continuity and Euler equation for the present wind
model are,
 \begin{align}
 \label{continuity}
 &{\pa \rho \over \pa t} + {\pa (\rho v)\over \pa z} = 0,\\
 \label{eulertimedep}
 &{\pa v \over \pa t} + v\, {\pa v \over \pa z} = -{z \over 1+z^2} +
\sqrt{2\rho\cri v\cri \; {\pa v /\pa z\over \rho}}.
 \end{align}
 In the Euler equation, $v'>0$ is assumed. The case $v'<0$ will be
discussed later. To derive Abbott waves, small perturbations are
applied, $v=v_0+v_1 \exp[i(kz-\omega t)]$ and $\rho=\rho_0+\rho_1
\exp[i(kz-\omega t)]$. The subscript 0 refers to a stationary
solution, subscript 1 to the perturbation. Linearizing gives
 \begin{gather}
 \label{continper}
 \left[-i {\omega \over v_0'}+i\chi +1 \right] {\rho_1 \over
\rho_0} + (i\chi -1) \; {v_1 \over v_0} =0,\\
 \label{eulerper}
 {1\over q_0} {\rho_1 \over \rho_0} + \left[-i {\omega \over v_0'}
+1 +i\chi \left(1-{1\over q_0}\right) \right] {v_1 \over v_0} =0,
 \end{gather}
 with non-dimensional $\chi\equiv k \,v_0/v_0'$ and $q_0=\sqrt{2 m_0
w_0'}$. Setting the determinant of the system to zero gives the
dispersion relation $\omega(k)$. In WKB approximation, $\chi\gg 1$,
one finds for the phase speed, $A_0$, and growth rate of the inward
($-$) mode,
 \begin{gather}
 \label{phasespeed}
 A_0 = {\rm Re}(\omega_-)/k = v_0 \left(1-{1\over q_0}\right),\\
       {\rm Im}(\omega_-)=0,\notag
 \end{gather}
 in the observers frame, and for the outward ($+$) mode,
 \begin{gather}
 \label{growthrate}
 A_0 = {\rm Re}(\omega_+)/k = v_0,\\
       {\rm Im}(\omega_+)=-\epsilon,\notag
 \end{gather}
 with $-\epsilon$ a small damping term of no further consequence.
Along a shallow solution, $q_0<1$, and along a steep solution,
$q_0>1$. For shallow solutions with small $m_0$ and $w_0'$, $q_0$ is
small, and the inward mode can propagate at arbitrarily large phase
speed, from {\it every} location $z$. This result is new, since so far
Abbott waves along shallow solutions were not considered (``a LDW has
no analog to a solar breeze,'' Abbott 1980). Since for the CAK point
star model every point is critical, $q_0=1$, and the inward mode is
stagnating everywhere. Stability of {\it non}-WKB waves, $\chi=O(1)$,
is found from a numerical solution.

Integrating (\ref{eulersol}), the velocity of the critical solution at
$z\cri=1$ (the `critical speed') is $v\approx 0.61$. On the other
hand, a shallow solution with $m_0=0.8$ has $v>0.61$ for $z>1.51$.
This is no contradiction to the fact that the shallow solution is
sub-abbottic.

\subsection{Characteristic analysis} 
\label{characteristicform}

The above analysis holds for linear waves, and the wave speed is
determined by the underlying stationary solution. General, non-linear
waves can be derived from a characteristic analysis. The
characteristic form of the above equations is, without further
approximations (Feldmeier \& Shlosman 2000),
 \begin{align}
 \label{charcon}
 &\left[{\pa \over \pa t} + v \, {\pa \over \pa z}\right] \rho = -\rho
          {\pa v\over \pa z},\\
 \label{chareul}
 &\left[{\pa \over \pa t} + A \, {\pa \over \pa z}\right] \left(
 {1\over\rho} {\pa v\over\pa z}\right) =- {1\over\rho} {\pa g\over\pa z},
 \end{align}
 where the Abbott speed, $A$, in the observers frame is
 \begin{equation}
 A=v \left(1-{1\over q}\right) = v-\sqrt{\rho\cri v\cri\over 2\rho\, \pa
v/\pa z},
 \end{equation}.

The interpretation of (\ref{chareul}) is unconventional, in
that $v'$ should now be considered as fundamental hydrodynamic field,
instead of $v$. This happens because of the nonlinear dependence of
$E$ on $v'$ (Courant \& Hilbert 1968).

Hence, $-\rho v'$ in the continuity equation and $-g'/\rho$ in the
Euler equation are inhomogenous terms, since they do not contain
derivatives of the hydrodynamic fields $\rho$ and $v'$. The advection
operators are $\pa/\pa t +v\,\pa/\pa z$ and $\pa/\pa t +A\,\pa/\pa z$,
with characteristic speeds $v$ and $A$. The former should be read as
$v+a$ in zero sound speed approximation. The Riemann invariants $\rho$
and $v'/\rho$ correspond to wave amplitudes. In a frame moving at $v$,
the amplitude $\rho$ is constant, except for sources and sinks $-\rho
v'$. In an isothermal gas, $p\sim\rho$, and the wave amplitude is
indeed that of a pressure wave. In a frame moving at $A$, the
amplitude $v'/\rho$ is constant, except for sources and sinks
$-g'/\rho$. The Sobolev optical depth is $\sim (v'/\rho)^{-1}$,
indicating that the ($-$) mode is a radiative wave.

Even for waves of infinitesimal amplitude, $v'$ may deviate strongly
from $v_0'$, if the wavelength $\lambda$ is short. Since $A\sim
(v')^{-1/2}$, short-scale Abbott waves are highly dispersive. For
sufficiently small $\lambda$, $v'$ becomes negative at certain wave
phases. The characteristic equations (\ref{charcon}, \ref{chareul})
hold also for $v'<0$, but $A$ has different meaning then. For a line
force $\sim \sqrt{|v'|}$, the general $A$ is
 \begin{equation}
 \label{outwardinward}
 A=v \left(1\mp {1\over q_\pm}\right), 
 \end{equation}
 where the lower sign applies for $v'<0$, and in $q_-$ one has to use
$-v'$. Trivially, Abbott waves are advected {\it outwards} in
super-abbottic flow. But that Abbott waves propagate {\it downstream}
as a $(+)$ mode in regions where $v'<0$ is probably their single, most
unique property.

For a line force $\sim \sqrt{\max(0,v')}$ then, $A=v$ if $v'<0$. This
is because the line force drops out of the Euler equation, and both
wave modes become ordinary sound. (Actually $A=v-a$ then, for upstream
propagating sound.) In this case one has along a characteristic with
coordinate $s$,
 \begin{equation}
 {\rho(s)\over\rho(0)} = {v'(s)\over v'(0)} = {1 \over 1+v'(0) s}.
 \end{equation}
 as WKB wave solution of (\ref{charcon}, \ref{chareul}). This
describes gas dynamical {\it simple} waves. Since $v'(0)<0$, they
steepen and break ($v'=-\infty$). The derivation of simple {\it
Abbott} waves is beyond the scope of this paper.

\section{Boundary conditions for line driven winds}
\label{boundaryconditions}

We discuss now the outer boundary conditions for shallow solutions.
Abbott waves can enter through the outer boundary, hence, pure outflow
boundary conditions are not appropriate for shallow solutions.

As hydrocode we use a standard Eulerian scheme on a staggered mesh.
Advection is solved in the conservative control volume approach, using
monotonic van Leer (1977) interpolation. The sound speed is set
exactly to zero, and a small amount of artificial viscosity is added
to handle occasional shock fronts. Details on the scheme can be found
in Reile \& Gehren (1991) and Feldmeier (1995).

In a first step, we consider simple boundary conditions for $\rho$ and
$v$ which still preserve essential features of Abbott waves. These
boundary conditions render shallow solutions stable, and allow small
perturbations to leave the grid. They are,
 \begin{gather}
 \label{simpleboundary}
 \rho(0)={\rm const}, \quad
 v(0)=v(1), \notag\\
 \rho(I)=\rho(I-1), \quad
 v(I)={\rm const},
 \end{gather}
 where 0 and $I$ are the mesh indices of the inner and outer boundary,
respectively. The conditions (\ref{simpleboundary}) are motivated as
follows. (The reader may jump to the end of this paragraph.) On a
characteristic curve $[z(s), t(s)]$ which leaves the mesh through a
boundary, the characteristic equation is solved on the boundary via
{\it extrapolation}, using one-sided, interior differentials. Such an
{\it upwinded} scheme is stable (Steger \& Warming 1981). For shallow
winds, the $\rho$ wave leaves through the outer boundary if $v>0$,
hence $\rho$ is extrapolated from the interior in
(\ref{simpleboundary}). Zeroth order extrapolation is used for
stability reasons. Abbott waves enter the mesh through the outer
boundary, hence a boundary condition must be applied. We choose
$v(I)={\rm const}$. (To prevent standing waves, $v(I)=m_0/\rho(I)$ is
often more appropriate; but is also more susceptible to runaway than
$v(I)={\rm const}$.)  At the {\it inner} boundary, the argument
proceeds correspondingly, with interchanged roles of the waves. We
find that using $\rho(0)=\rho(1)$ and $v(0)={\rm const}$ instead
destabilizes shallow solutions.

\subsection{The Courant time step} 
\label{runawayI}

In time-dependent hydrodynamic simulations published so far, it is
customary to insert the {\it sound} speed in the Courant time step,
 \begin{equation}
 \label{courant}
 \Delta t = \sigma \min \;\; {\Delta z \over |v \pm a|},
 \end{equation}
 with Courant number $\sigma\le 1$, and the minimum has to be taken
over the mesh. We see now that Abbott waves, as characteristic inward
wave mode, have to be included in the Courant time step,
 \begin{equation}
 \Delta t = \sigma \min \;\; \left(
 {\Delta z \over |v-A|},  {\Delta z \over v+a} \right),
 \end{equation}
 where $A$ is in the fluid frame now. Note that $A$ changes sign with
$v'$. For sufficiently small $v'>0$ and for arbitrary $v'<0$, Abbott
waves determine the time step. At velocity plateaus where the wind
velocity is more or less constant, $A\rightarrow -\infty$ from
(\ref{chareul}), and the Courant time step $\rightarrow 0$. This is an
artifact of Sobolev approximation in which the line radiation force
tends to zero in the absence of velocity gradients. In practice, in
calculating the Courant time step we do not allow $q$ to drop below a
certain minimum value, usually $10^{-4}$. This corresponds to $\sigma
\approx 10^{-4}$ for an ordinary Courant time step not including
Abbott waves. Our results are not sensitive to the value of this
minimum $q$, if the latter is sufficiently small. Furthermore,
velocity plateaus quickly acquire a tilt and propagate at finite
speed.

\subsection{Non-reflecting boundary conditions} 
\label{riemann}

We can then formulate {\it non-reflecting} boundary conditions for the
Riemann invariants $\rho$ and $v'/\rho$, which annihilate incoming
waves. This prevents boundary reflection of waves which originate on
the interior mesh. To annihilate a wave, its amplitude (the Riemann
invariant) is kept constant in time at the boundary, $\pa(v'/\rho)/\pa
t=0$ and $\pa\rho/\pa t=0$ for Abbott and sound waves, respectively
(Hedstrom 1979; Thompson 1987). One may say that non-reflecting
boundary conditions drive the numerical solution towards a
neighboring, stationary solution without waves. The technical details
of implementing these boundary conditions are discussed in Appendix~A.

\section{Stability of solutions} 
\label{stability}

We discuss here the numerical stability of shallow and steep
solutions, depending on the appropriate Courant time step and boundary
conditions.

\subsection{Stability of steep solutions}

We start with steep solutions, and show that they are unconditionally
stable, even when simplified boundary conditions analogous to
(\ref{simpleboundary}) are used. According to (\ref{eulersol}), $q>1$
everywhere for steep solutions, and Abbott waves can propagate only
towards larger $z$. At the outer boundary, extrapolation
$\rho(I)=\rho(I-1)$ and $v(I)=v(I-1)$ is therefore appropriate. At the
inner boundary, two conditions have to be specified. The obvious
choices are either $\rho={\rm const}$, $v={\rm const}$ or $\rho={\rm
const}$, $v'={\rm const}$. The first set fixes the mass loss rate, the
second establishes non-reflecting boundary conditions: the wave
amplitudes are $\rho$ and $v'/\rho$, and the latter boundary
conditions keep them constant in time, meaning there are no waves.

We find that steep solutions are stable for both types of boundary
conditions. Even when the initial conditions differ profoundly from a
steep wind, the numerical code converges to the latter. Furthermore,
we performed tests with an explicit, harmonic perturbation at fixed
Eulerian position in the wind. Even if the perturbation amplitude
reached 100\%, the wind remained on average on a steep solution, which
can, therefore, be considered as unconditionally stable.

\subsection{Stability and runaway of shallow solutions}

A physically more relevant question is for the stability of shallow
solutions. In a first step, we use an analytic, shallow wind as
initial conditions. Table~\ref{table1} shows the {\it runaway} of this
shallow to the critical velocity law. Abbott waves are {\it not}
accounted for in the Courant time step, and pure outflow boundary
conditions $\rho(I)=\rho(I-1)$ and $v(I)=v(I-1)$ are applied. This
corresponds to the typical procedure adopted in the literature. The
runaway starts at the outer boundary and generates inward propagating
Abbott waves. The velocity law evolves towards larger speeds, until
the critical solution is reached.

 \begin{table}
 \begin{center}
 \caption{\label{table1} Numerical runaway of a shallow velocity law.}
 \smallskip
 \begin{tabular}{c|rrrrrr}
 \hline\hline\smallskip
 $\downarrow z, t\rightarrow$&0&3&6&9&20&$\infty$\cr
 \hline
 0.1&{\bf 0.0}&{\bf 0.0}&{\bf 0.0}&{\bf 0.0}&{\bf 0.0}&{\bf 0.0}\cr
 0.5&{\bf 0.2}&{\bf 0.2}&{\bf 0.2}&{\bf 0.2}&{\bf 0.2}&{\bf 0.2}\cr
 1.0&     0.5 &{    0.4}&{\bf 0.6}&{\bf 0.6}&{\bf 0.6}&{\bf 0.6}\cr
 1.5&     0.6 &{    0.6}&{    0.9}&{\bf 1.0}&{\bf 1.0}&{\bf 1.0}\cr
 2.0&{\it 0.7}&{    0.8}&{\it 1.2}&{    1.4}&{\bf 1.4}&{\bf 1.4}\cr
 2.5&{\it 0.8}&{    0.9}&{\it 1.3}&{\it 1.5}&{\bf 1.8}&{\bf 1.8}\cr
 3.0&{\it 0.8}&{\it 1.1}&{\it 1.3}&{\it 1.5}&{\bf 2.1}&{\bf 2.1}\cr
 3.5&{\it 0.8}&{\it 1.1}&{\it 1.4}&{\it 1.6}&{\it 2.2}&{\bf 2.4}\cr
 4.0&{\it 0.8}&{\it 1.2}&{\it 1.4}&{\it 1.6}&{\it 2.3}&{\bf 2.7}\cr
 \end{tabular}
 \end{center}
 \end{table}

The mechanism of the runaway seems as follows. The Courant condition
for Abbott waves is violated in the outer wind, where the velocity law
is flat and the Abbott speed is large. This causes numerical
runaway. Opposed to standard gas dynamics, the LDW runaway does not
crash the simulation. We speculate that exponential growth in $v$
creates large $|v'|$, which imply small Abbott speeds. The Courant
condition is no longer violated, and runaway stops. Abbott waves are
also excited at the outer boundary, by switching from the interior to
the boundary scheme. Upstream propagating Abbott waves are
inconsistent with the assumed outflow boundary conditions. The waves
propagate to the wind base and drive the inner wind towards the
critical solution, which is indeed consistent with outflow boundary
conditions.

Note in Table~\ref{table1} that Abbott waves can propagate inwards
from the outer boundary, $\zM=4$, until the critical velocity law is
reached (indicated by bold numbers in the table). This is no
contradiction to the critical point being located at $z\cri=1$: the
outer, shallow portion of the velocity law (the numbers in italics) is
sub-abbottic, even when the inner, steep velocity law is already
super-abbottic.

To {\it prevent numerical runaway} of shallow solutions, we apply
non-reflecting Riemann boundary conditions and include Abbott waves in
the Courant time step. As initial conditions, an arbitrary (for
example, linear) velocity law is used. Abbott waves are again exited
at the outer boundary and propagate inwards. However, the simulation
relaxes quickly to a shallow solution. The $m$-value depends on the
initial data. Even for an initial model with height-dependent mass
flux, we find quick convergence to a shallow solution.

Further details on the stability of shallow solutions are given in
Appendix~B. To summarize this section, both improper outer boundary
conditions and Courant conditions which do not account for Abbott
waves are responsible for numerical runaway.

\section{Summary}
\label{summary}

We present a simplified model for line driven winds from stars and
accretion disks which avoids the $r^{-2}$ degeneracy of the CAK
model. Radiative or Abbott waves are derived, both from a dispersion
and a characteristic analysis, and for all possible solutions to the
Euler equation, i.e., shallow, critical, steep and overloaded ones.

We find that Abbott waves can propagate upstream, towards the
photosphere, from any position along a shallow wind solution. Hence,
for the latter, they can enter the calculational grid through the
outer boundary. The wave propagation speed depends on the velocity
slope of the wind. Abbott's (1980) result that these waves are
creeping, i.e., have only slighly negative inward speeds, holds
globally only for the almost degenerate CAK point-star model. In more
realistic wind models, Abbott waves can limit the Courant time step.

Both the neglect of Abbott waves in the Courant time step and in the
boundary conditions leads to numerical runaway, towards the critical
wind solution of maximum mass loss rate or to accretion flow. If,
instead, incoming waves are annihilated at the outer boundary and the
correct Courant timestep is used, shallow solutions are stable.

\acknowledgements

It is a pleasure to thank Janet Drew, Leon Lucy, Stan Owocki, and
Joachim Puls for frequent discussions. This work was supported in part
by NASA grants NAG 5-10823, NAG5-3841, WKU-522762-98-6 and HST
GO-08123.01-97A to I.S., which are gratefully acknowledged.

\appendix

\section{Non-reflecting boundary conditions}

We adopt the following procedure to calculate $\rho$ and $v$ on a
boundary. (i) If the $\rho$ wave leaves the grid, $\dot\rho= -(\rho
v)'$ is calculated using one-sided, interior differentials. (ii) If
the $\rho$ wave enters the grid, $\dot \rho=0$ is set (wave
annihilation; see the main text). (iii) If the $v'/\rho$ wave leaves
the grid, $\dot v'/v'$ is calculated from the Euler
equation~(\ref{chareul}) in characteristic form, using one-sided,
interior differentials. Note that $v''$ appears here. (iv) If
$v'/\rho$ enters the grid, $\dot v'/v'=\dot \rho/\rho$ is set,
assuming WKB waves. (v) The new values for $\rho$ and $v$ on the
boundary are found by integrating $\dot\rho$ and $\dot v'$ using a
time-explicit scheme.

In the latter step (v), one actually needs $\dot v$, not $\dot v'$.
To integrate from time step $n-1$ to $n$ at the outer boundary, using
$\dot v'$, we write
 \begin{equation}
 {v'}^n = {v'}^{n-1} + \dot v' \Delta t.
 \end{equation}
 Using ${v'}^n =(v^n_I - v^n_{I-1})/\Delta z$, and similarly for
${v'}^{n-1}$, this becomes,
 \begin{equation}
 \label{memory}
 v^n_I = v^{n-1}_I + v^n_{I-1} - v^{n-1}_{I-1} + \dot v' \Delta
z\Delta t,
 \end{equation}
 whence $v_n^I$ depends on $v^{n-1}_{I-1}$. Due to frequent variable
updating in an operator splitting scheme during the time step,
$v^{n-1}_{I-1}$ would have to be stored as an extra variable. This
seems not desirable, and causes indeed numerical runaway of shallow
solutions. Fortunately, a simple approximation to (\ref{memory}) gives
satisfactory results, namely setting $v^{n-1}_{I-1} \equiv
v^{n}_{I-1}$, or
 \begin{equation}
 \label{boundaryscheme}
 v^n_I = v^{n-1}_I + \dot v' \Delta z\Delta t.
 \end{equation}

\section{Numerical stability of shallow solutions}

For numerical stability tests, we assume the following parameters, if
not otherwise stated: 200 equidistant grid points from $z=0.1$ to
$z=4$. Below $z=0.1$, the stationary wind speed is very small. Since
$\delta v/ A_0 =\delta \rho /\rho_0$, one has $v_0\ll A_0$ for an
inner boundary $\zm\rightarrow 0$. Even small perturbations $\delta
\rho$ cause then negative speeds at $\zm$. This alters the direction
of the $\rho$ characteristic, and often causes numerical problems. A
shallow solution with $m=0.8$ serves as start model. The sound speed
is set to zero. The flow time from $z=0.1$ to $z=4$ is 12.4 for a
shallow solution with $m=0.8$, and 9.1 for the critical solution (in
the units specified in the main text). For $m=0.8$, linear Abbott
waves propagate from $z=4$ to $z=0.1$ in a time 6.0. All simulations
are evolved to time 50. Table~\ref{tableapp} lists results from
simulations using different outer boundary conditions and Courant time
steps. On the inner boundary, conditions (\ref{simpleboundary}) were
used.

 \begin{deluxetable}{cccccccc}
 \tablecolumns{8}
 \tablewidth{33pc}
 \tablecaption{\label{tableapp} Runaway in simulations applying
different boundary conditions and Courant time steps.}
 \tablehead{
 \colhead{run} & \colhead{$\zM$} & \colhead{grid} &
\colhead{line} & \colhead{Abbott} & \colhead{Courant} &
\colhead{boundary} & \colhead{result}\\ 
 \colhead{} & \colhead{} & \colhead{points} & \colhead{force} &
\colhead{time} & \colhead{number} & \colhead{condition} & \colhead{}}
 \startdata
 1 &  4 & 200 & abs$(v')$   & no  & 0.5 & outflow & runaway           \nl
 2 &  4 & 200 & abs$(v')$   & no  & 0.1 & outflow & $m_0=0.7$; oscill.\nl
 3 &  4 & 200 & abs$(v')$   & no  & 0.5 & shallow & runaway           \nl
 4 &  4 & 200 & abs$(v')$   & no  & 0.1 & shallow & runaway           \nl
 5 & 10 & 200 & abs$(v')$   & no  & 0.1 & outflow & runaway           \nl
 6 &  4 & 200 & abs$(v')$   & yes & 0.5 & outflow & $m_0=0.5$; oscill.\nl
 7 & 10 & 200 & abs$(v')$   & yes & 0.5 & outflow & slow runaway      \nl
 8 & 10 & 500 & abs$(v')$   & yes & 0.5 & outflow & slow runaway      \nl
 9 &  4 &  80 & abs$(v')$   & yes & 0.5 & outflow & $m_0=0.5$; oscill.\nl
10 &  4 & 200 & abs$(v')$   & yes & 0.5 & shallow & stable            \nl
		    
11 &  4 & 200 & max$(0,v')$ & no  & 0.5 & outflow & accretion         \nl
12 &  4 & 200 & max$(0,v')$ & no  & 0.1 & outflow & accretion         \nl
13 &  4 & 200 & max$(0,v')$ & no  & 0.5 & shallow & stable, oscill.   \nl
14 &  4 & 200 & max$(0,v')$ & no  & 0.1 & shallow & stable, oscill.   \nl
15 & 10 & 200 & max$(0,v')$ & no  & 0.1 & outflow & accretion         \nl
16 &  4 & 200 & max$(0,v')$ & yes & 0.5 & outflow & accretion         \nl
17 & 10 & 200 & max$(0,v')$ & yes & 0.5 & outflow & accretion         \nl
18 & 10 & 500 & max$(0,v')$ & yes & 0.5 & outflow & accretion         \nl
19 &  4 &  80 & max$(0,v')$ & yes & 0.5 & outflow & accretion         \nl
20 &  4 & 200 & max$(0,v')$ & yes & 0.5 & shallow & stable            \nl
\enddata
\end{deluxetable}

Consider first the line force $\sim \sqrt{|v'|}$. The table shows that
inclusion of Abbott waves in the Courant time is mandatory to prevent
runaway (runs~1-5; `shallow' in the column for the outer boundary
conditions refers to (\ref{simpleboundary}). For small Courant
numbers, Abbott waves are, to a degree, already accounted for by an
ordinary time step. Runaway may then be prevented, as in run~2, though
the solution does oscillate at all times. Shifting the outer boundary
outwards, the runaway occurs again (run~5). If Abbott waves are
accounted for in the time step, outflow boundary conditions
(extrapolation of $\rho$ and $v$) do not necessarily lead to
runaway. Instead, a stable, shallow solution may be maintained
(runs~6, 9; also 2). However, the solution oscillates at all times.
Shallow solutions become more unstable when the outer boundary is
shifted outwards, into the shallow part of the velocity curve where
Abbott waves are easily excited. Finally, the control run~10 shows
that the initial shallow solution is numerically stable if Abbott
waves are included in the time step and boundary conditions
(\ref{simpleboundary}) are used.

For a line force $\sim \sqrt{\max(0,v')}$, the scheme is even more
susceptible to numerical runaway. All cases with outflow boundary
conditions undergo runaway. However, for the boundary condition
$v(I)=v(I-1)$, the solution no longer tends towards the critical
solution. Instead $v(I)$ drops to zero, and with it the interior wind
velocity. The outcome of the simulation is not clear, as we have not
formulated appropriate line driven {\it accretion} boundary
conditions. The reason for the drop in $v(I)$ is that an accidental
$v'<0$ at the outer boundary implies zero line force. This causes
further velocity drop and $v'$ becomes more negative. Abbott waves
carry this wind breakdown to the interior mesh. It is not clear
whether this type of line force should actually be applied in the
proximity of the outer boundary.


\begin{references}

\reference{} Abbott, D. C. 1980, ApJ, 242, 1183
\reference{} Abbott, D. C. 1982, ApJ, 259, 282
\reference{} Castor, J. I., Abbott, D. C., \& Klein, R. I. 1975, ApJ, 195, 157
\reference{} Courant, R., \& Hilbert, D. 1968, Methoden der
             Mathematischen Physik, Springer, Berlin
\reference{} Cranmer, S. R., \& Owocki, S. P. 1996, ApJ, 462, 469
\reference{} Feldmeier, A. 1995, A\&A, 299, 523
\reference{} Feldmeier, A., \& Shlosman, I. 1999, ApJ, 526, 344
\reference{} Feldmeier, A., \& Shlosman, I. 2000, ApJ, 532, L125
\reference{} Hedstrom, G. W. 1979, J. Comp. Phys., 30, 222
\reference{} Kudritzki, R. P., Puls, J., Lennon, D. J., et al. 1999, A\&A, 350, 970
\reference{} Lamers, H. J. G. L. M., \& Leitherer, C. 1993, ApJ, 412, 771
\reference{} Lamers, H. J. G. L. M., \& Cassinelli, J. P. 1999, Introduction
             to Stellar Winds (Cambridge: Cambridge Univ. Press) 
\reference{} Owocki, S. P., \& Puls, J. 1999, ApJ, 510, 355
\reference{} Owocki, S. P., \& Rybicki, G. B. 1986, ApJ, 309, 127
\reference{} Pauldrach, A. 1987, A\&A, 183, 295
\reference{} Poe, C. H., Owocki, S. P., \& Castor, J. I. 1990, ApJ, 358, 199
\reference{} Puls, J., Kudritzki, R. P., Herrero, A., et al. 1996, A\&A, 305, 171
\reference{} Puls, J., Springmann, U., \& Lennon, M. 2000, A\&AS, 141, 23
\reference{} Reile, C., \& Gehren, T. 1991, A\&A, 242, 142
\reference{} Steger J. L., \& Warming R. F. 1981, J. Comp. Phys., 40, 263
\reference{} Thompson, K. W. 1987, J. Comp. Phys., 68, 1
\reference{} van Leer, B. 1977, J. Comp. Phys., 23, 276

\end{references}
\end{document}